\pdfoutput=1
\pdfoutput=1
\pdfoutput=1
\pdfoutput=1
\pdfoutput=1
\documentclass[nofootinbib,showpacs,eqsecnum]{revtex4}

\usepackage[dvips]{color}
\usepackage{epsfig}
\usepackage{amsmath}
\usepackage{graphicx}

\begin{document}
\title{Thermodynamics and holography of tachyon cosmology}
\author{H. Farajollahi$^{1,2}$}
\email{hosseinf@guilan.ac.ir} \author{A. Ravanpak$^{1}$}\email{aravanpak@guilan.ac.ir}\author{M. Abolghasemi$^{1}$}\email{masood@msc.guilan.ac.ir}
\affiliation{$^1$Department of Physics, University of Guilan, Rasht, Iran}
\affiliation{$^2$ School of Physics, University of New South Wales, Sydney, NSW, 2052, Australia}

\date{\small {\today}}

\def\be{\begin{equation}}
\def\ee{\end{equation}}
\def\bea{\begin{eqnarray}}
\def\eea{\end{eqnarray}}
\def\M{{{\cal M}}}
\def\bdy{{\partial\cal M}}
\def\w{\widehat}
\def\n{\widetilde}
\def\real{{\bf R}}

\begin{abstract}
Recently, we have investigated the dynamics of the universe in tachyon cosmology with non-minimal coupling to matter \cite{faraj}-\cite{faraj3}. In particular, for the interacting holographic dark energy (IHDE), the model is studied in \cite{Ravanpak}. In the current work, a significant observational program has been conducted to unveil the model's thermodynamic properties. Our result shows that the IHDE version of our model better fits the observational data than $\Lambda$CDM model. The first and generalized second thermodynamics laws for the universe enveloped by cosmological apparent and event horizon are revisited. From the results, both first and generalized second laws, constrained by the observational data, are satisfied on cosmological apparent horizon.In addition, the total entropy is verified with the observation only if the horizon of the universe is taken as apparent horizon. Then, due to validity of generalized second law, the current cosmic acceleration is also predicted.

\end{abstract}
\pacs{98.80.Cq; 11.25.-w}

\keywords{non-minimal coupling; tachyon; interacting; holographic; thermodynamics}
\maketitle

\newpage
\section{INTRODUCTION}

Various cosmological observations, including the Type Ia Supernova (Sne Ia) \cite{Riess},\cite{Perlmutter}, the cosmic microwave background radiation \cite{Spergel},\cite{Spergel2} and the large scale structure \cite{Tegmark},\cite{Eisenstein}, have revealed that the universe is in accelerating state. The most appealing candidate to explain this phenomenon is dark energy (DE) that in a simplest form as the vacuum energy ( i.e. cosmological constant, $\Lambda$) inherits a constant equation of state (EoS) parameter, $w = -1$. Further, the DE models with time dependent EoS parameter have also been the subject of extensive investigation for nearly one decade. For example, cosmological models with quintessence \cite{Caldwell}, phantom \cite{Caldwell2}, k-essence \cite{Armendariz}, tachyon \cite{Padmanabhan2},\cite{Sen} and quintom fields \cite{Feng},\cite{eli}; as well as the Chaplygin gas \cite{Kamenshchik}, the generalized Chaplygin gas \cite{Bento}, holographic DE (HDE)\cite{Cohen},\cite{Li}, old and new agegraphic DE \cite{Wei},\cite{Wei2}, and the Ricci DE \cite{Gao} are among the most promising models to reveal the physical phenomena in the universe dynamics.

From the above candidates for DE the HDE model is an attempt to apply the nature of DE within the framework of quantum gravity \cite{Hsu} that its energy density is given by,
\begin{equation}\label{rol}
\rho_\Lambda = 3c^2M_p^2/L^2.
\end{equation}
In (\ref{rol}), $M_P$ is Planck mass, $c$ is a free dimensionless parameter with a constraint given by $c \geq 1$ \cite{Huang} and coefficient three is for convenience. From \cite{Li} we find that $L$ should be the future event horizon,
\begin{equation}\label{Re}
R_E=a\int_a^\infty\frac{da'}{Ha'^2},
\end{equation}
with $a$ representing scale factor of the universe \cite{Wang}.

The interacting HDE model (IHDE), is a HDE model which includes an interaction between dark sectors. It has been argued that with the interaction between dark energy and dark matter the old astrophysical structures can be formed naturally \cite{Bin} and the coincidence problem can be alleviated \cite{Bin}--\cite{Sand}.

Also, among scalar field theories, both non-minimal coupling of the field to matter and tachyon model have been widely investigated in last few years \cite{mohseni2}--\cite{wung} and relevant works in the holographic tachyon models in both interacting and non interacting cases can be found in \cite{Bagla}--\cite{Setare5}.

On the other hand, the relations between thermodynamics of space time and the nature of
gravity is one of the most intriguing topics in theoretical physics. The
thermodynamics laws of black hole were first derived from the classical Einstein
equation \cite{carter} and followed by discovery of Hawking radiation of black holes
\cite{Hawking}, \cite{Bekenstein}. Separately it has been shown that from the proportionality
of entropy and horizon area together with the first law of thermodynamics one can derive the Einstein equation \cite{Jacobson}. This implies a close relation between thermodynamics
of space time and the Einstein equation \cite{Verlinde}, \cite{Padmanabhan}. In \cite{Cai}, the authors, by applying the first law of thermodynamics to the apparent horizon and assuming the geometric entropy is given by a quarter of the apparent horizon area, study the FRW cosmology.

Also, the thermodynamical properties of several cosmological models have been investigated recently and the authors have probed the first and second laws of thermodynamics in them. In \cite{Bhattacharya} the validity of thermodynamical laws in a dark energy filled universe has been discussed and in \cite{Bamba}--\cite{Bamba2} the laws of thermodynamics in modified gravity theories have been investigated. In \cite{Chattopadhyay} the author deals with the second law of thermodynamics in the presence of interacting tachyonic field and another scalar field. Also, in some HDE and IHDE models the authors have tried to discuss the laws of thermodynamics in flat and non-flat universes \cite{Mazumder}--\cite{Setare2}. In here, extending the previous works, we would like to study the thermodynamics properties of an IHDE, by considering a tachyon cosmological model non-minimally coupled to matter Lagrangian.

\section{The Model and Cosmological constraints }

We start with the action
\be\label{action}
S=\int{[\frac{R}{16\pi}-V(\phi)\sqrt{1-\dot{\phi}^2}+f(\phi){\cal{L}}_m]\sqrt{-g}d^4x}
\ee
where $R$ and ${\cal{L}}_m$ are Ricci scalar and matter Lagrangian respectively, $\phi$ is a scalar tachyon field and $V(\phi)$ represents its potential and $f(\phi)$ is a scalar function where indicates the non-minimal coupling between matter and tachyon field. In a spatially flat FRW  cosmology, the Friedmann equation is given by \cite{Ravanpak}
\begin{equation}
H^{2}=\frac{8\pi}{3}(\rho_{m}f(\phi)+\frac{V(\phi)}{\sqrt{1-\dot{\phi}^{2}}}),\label{fried1}\\
\end{equation}
where $ H=\frac{\dot{a}}{a}$. Considering a HDE and an interaction between DE and cold dark matter (CDM), the dynamical equation for DE is given by
\be\label{Omega_tac}
\frac{d\Omega_{tac}}{dz}=-(1+z)^{-1}\Omega_{tac}[(1-\Omega_{tac})(1+\frac{2\sqrt{\Omega_{tac}}}{c})-\frac{8\pi Q}{3H^{3}}],
\ee
where $Q=\epsilon\rho_m\dot f$ and $\epsilon=\frac{1-3\gamma}{4}$ where $\gamma$ is the EoS parameter of the matter in the universe.
Also, from \cite{Ravanpak} we have the evolution of tachyon field and Hubble parameter as
\be\label{phiprime}
\frac{d\phi}{dz}=\frac{\sqrt{1+w_{tac}}}{H(1+z)},
\ee
\begin{equation}\label{Hz3}
\frac{dH}{dz}= -(1+z)^{-1}H(-\frac{3}{2}+\frac{\Omega_{tac}}{2}+\frac{\Omega_{tac}^{3/2}}{c}+\frac{4\pi Q}{3H^3}),
\end{equation}
where the EoS parameter for tachyon field representing DE is given by
\be\label{omega_tac}
w_{tac}=-\frac{1}{3}-\frac{2\sqrt{\Omega_{tac}}}{3c}-\frac{8\pi Q}{9H^{3}\Omega_{tac}}\cdot
\ee
From the above equations one can easily obtain the total EoS and deceleration parameters $q$ for the model as
\begin{eqnarray}
w_{tot}&=&-\frac{2}{3c}\Omega_{tac}^{3/2}-\frac{1}{3}\Omega_{tac}-\frac{8\pi Q}{9H^3},\label{w}\\
q&=&-\frac{\Omega_{tac}^{3/2}}{c}-\frac{1}{2}\Omega_{tac}+\frac{1}{2}-\frac{4\pi Q}{3H^3}.\label{q}
\end{eqnarray}
In the following, we assume an exponential form of $f(\phi)=f_0e^{b\phi(z)}$ where the interacting parameters $b$ and $f_0$ together with $\rho_m$ and $\gamma = 0$ for CDM determine the dynamics of the system. Obviously, from interacting term $Q$, the $f_0=0$ or $b=0$ leads to the absence of the interaction. Also, considering $\gamma=1/3$ for radiation leads to $\epsilon=0$ and again we will not have any interaction. But in general because of the smallness of the interacting term $Q$, different values of $\gamma$ does not affect the following results.

Next, we find the constraints on the model parameters using $\chi^2$ method by utilizing recent observational data, including the observational Hubble data (OHD), the baryonic acoustic oscillation (BAO) distance ratio and the cosmic microwave background (CMB) radiation. By solving the equations (\ref{Omega_tac})-(\ref{Hz3}) for the dynamical variables $\Omega_{tac}$, $\phi$ and $H(0)$ and the parameters $c$ and $b$, the constraints from a combination of OHD, BAO and CMB can be obtained by minimizing $\chi^2_{OHD}+\chi^2_{BAO}+\chi^2_{CMB}$. The results have been shown in Table \ref{table:1}. For comparison, we also constrained the parameters $H(0)$ and  $\Omega_{m0}$ for $\Lambda$CDM model. Table \ref{table:1} shows that, for OHD data, the $\Lambda$CDM model fits the observational data better than IHDE model. However, considering the combination of OHD + BAO + CMB, our model better fits the data in comparison to $\Lambda$CDM model. This can be justified by the fact that the contribution of BAO and  CMB data to the model are from earlier stage of universe whereas $\Lambda$CDM model describes only the recent behavior of the universe.

\begin{table}[ht]
\caption{Constraints on model parameters and initial conditions in our model and $\Lambda$CDM model} 
\centering 
\begin{tabular}{|c|c|c|c|c|c|c|c|c|c|} 
\hline\hline 
-& parameters & H(0) & $\Omega_{m0}$ &$\Omega_{tac}(0)$& $\phi(0)$ & c & b & $\chi^2_{min}$ & $\chi^2_{red}=\chi^2_{min}/dof$\\ [3ex] 
\hline\hline 
 Model &OHD & 0.71 &-& 0.759 & 1.1& 1 & -0.6 & 7.941 & 0.882 \\
\hline 
 Model & OHD + BAO + CMB & 0.71 &-& 0.750 & 0.6& 1.01 & 1.3 & 9.467 & 0.789 \\
\hline 
$\Lambda$CDM  & OHD   & 0.71 &0.29& - & -& - & - & 7.954 & 0.663 \\
\hline 
$\Lambda$CDM  & OHD + BAO + CMB & 0.61 &0.63& -& -& - & - & 20.382 & 1.359 \\
\hline
\end{tabular}
\label{table:1} 
\end{table}\

\begin{figure}[h]
\centering
\includegraphics[width=0.3\textwidth]{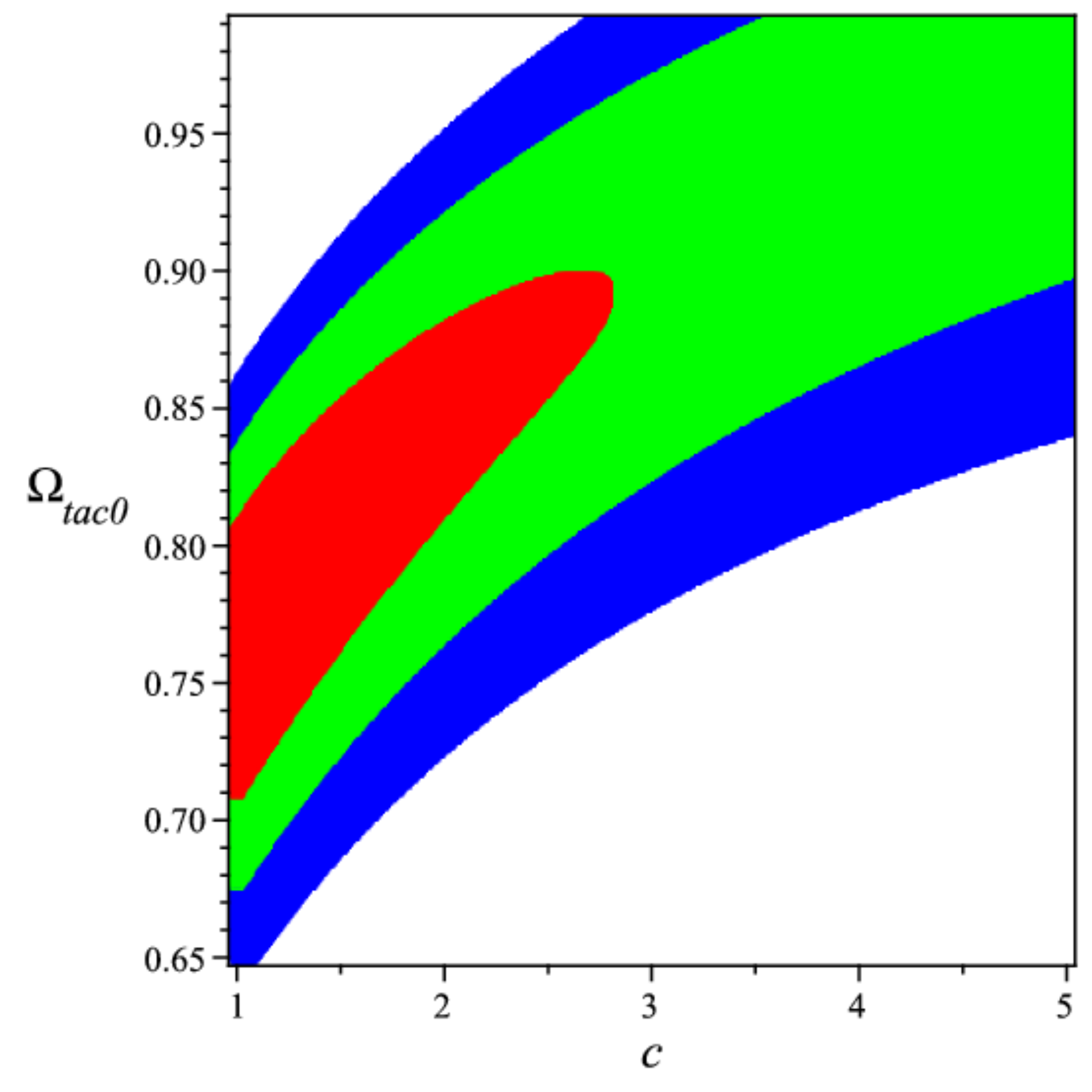}
\includegraphics[width=0.3\textwidth]{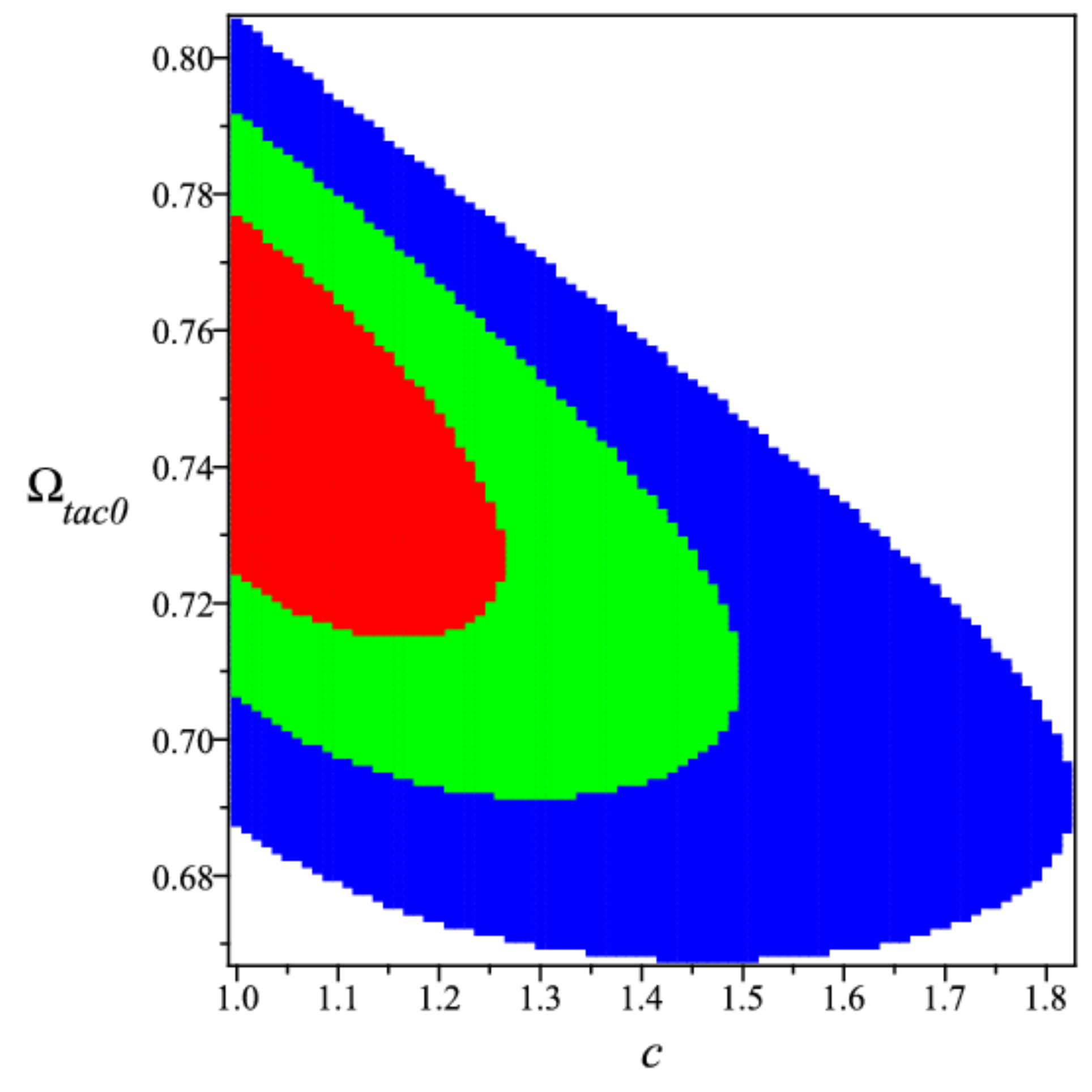}
\caption{The contour plots of best fitted parameters $\Omega_{tac}(0)$ and $c$ using left) OHD data, right) combined OHD + BAO + CMB data. The red, green and blue regions show 68.3\%, 95.4\% and 99.7\% confidence regions, respectively.}\label{conf}
\end{figure}

For example, the contour plots have been presented in Fig.\ref{conf} show the confidence regions for the best fitted $\Omega_{tac}(0)$ and  $c$ in our model using OHD and combined OHD + BAO + CMB data, respectively. We learn from this figure and also Table \ref{table:1} that the combination of observational data make stronger constraints on this model parameters.

\section{THERMODYNAMICAL ANALYSIS}

By ascribing HDE to the tachyon field in the model, $8\pi\rho_{tac}=3\Omega_{tac}H^2=24\pi c^2R_{E}^{-2}$, the event horizon becomes $R_{E}=c\sqrt{8\pi}/(\sqrt{\Omega_{tac}}H)$. With the apparent horizon given by $R_{A}=H^{-1}$, the ratio between apparent and event horizon becomes $\frac{R_{A}}{R_{E}}=\sqrt{\Omega_{tac}}/(c\sqrt{8\pi})$. So, one easily find that the apparent horizon is in general smaller than the event horizon.

\begin{figure}
\centering
\includegraphics[width=0.4\textwidth]{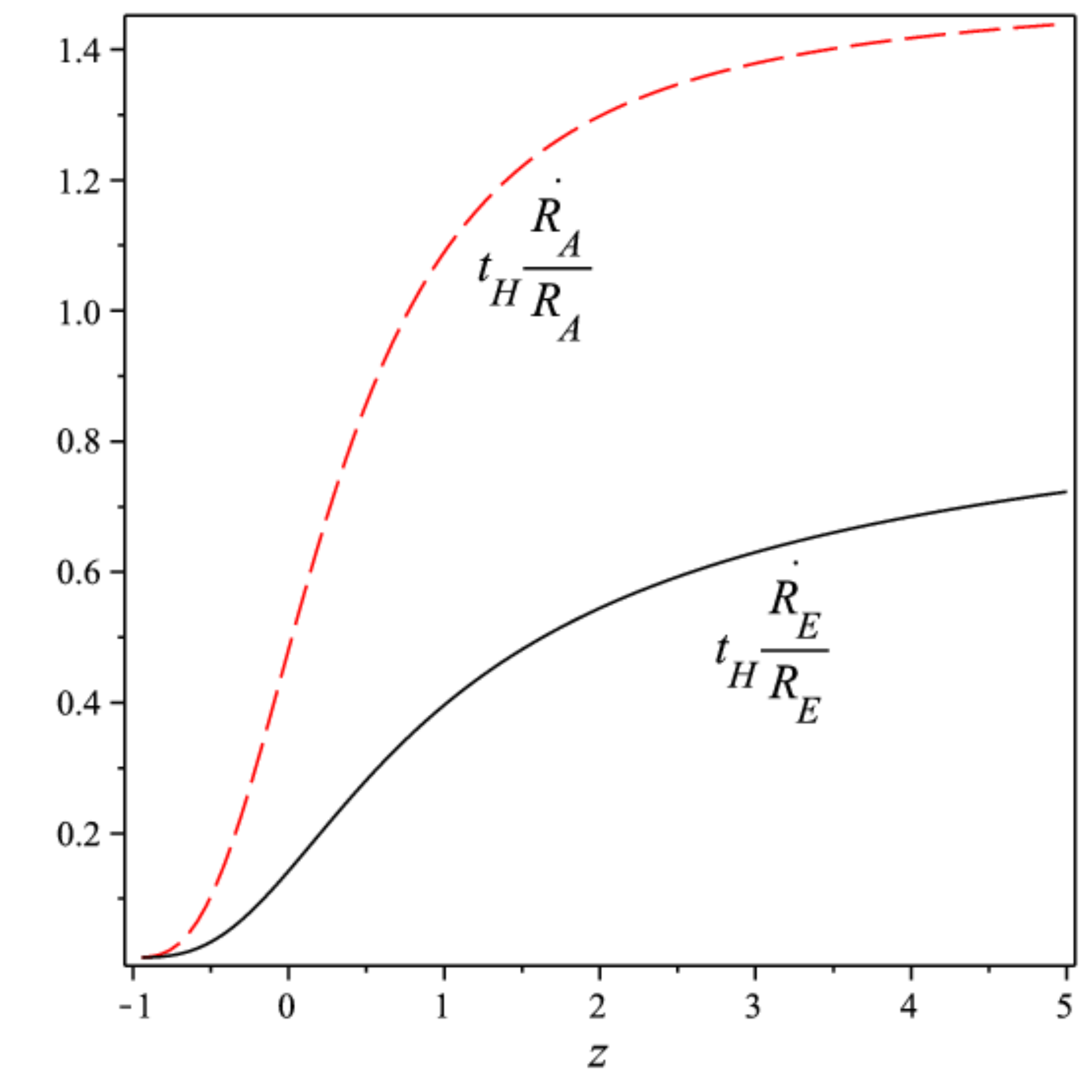}
\caption{The rate changes of the apparent horizon (red dashed curve) and the event horizon (black solid curve) over one Hubble scale. For plotting we have used the best fit values.}\label{tHtH}
\end{figure}

As the thermodynamic laws are applicable in equilibrium, we first investigate whether these distances change dominantly over one hubble time and then we study the validity of the thermodynamic laws. So, let us calculate how much the horizons change over one Hubble scale. For the apparent horizon, we have
\be\label{tH}
t_H\frac{\dot{R_{A}}}{R_{A}}=\frac{(1+z)}{H}\frac{dH}{dz}\cdot
\ee
So using (\ref{Hz3}), we obtain
\bea\label{tHfinal}
t_{H}\frac{\dot{R_{A}}}{R_{A}}&=&\frac{3}{2}-\frac{\Omega_{tac}}{2}-\frac{\Omega^{3/2}_{tac}}{c}-\frac{4\pi Q}{3H^{3}}
\eea
On the other hand, for the event horizon, we have
\be\label{tH2}
t_H\frac{\dot{R_{E}}}{R_{E}}=\frac{(1+z)}{H}\frac{dH}{dz}+\frac{(1+z)}{2\Omega_{tac}}\frac{d\Omega_{tac}}{dz}\cdot
\ee
Similar to the case of apparent horizon, using differential equation for DE, (\ref{Omega_tac}) and with attention to (\ref{Hz3}) , we find
\bea\label{tH2final}
t_{H}\frac{\dot{R_{E}}}{R_{E}}&=&1-\frac{\sqrt{\Omega_{tac}}}{c}
\eea
 From numerical calculation and utilizing the best fitted model parameters, in Fig.(\ref{tHtH}), we see that neither the apparent horizon nor the event horizon significantly change over one Hubble scale. Thus, the equilibrium thermodynamics can be applied here. Resembling the black hole case, we have
\begin{eqnarray}\label{blackhole}
    S_X=\pi R_X^2& , &T_X=1/(2\pi R_X)
\end{eqnarray}
where $X$ stands for apparent ($A$) or event ($E$).

To investigate the first law of thermodynamics on apparent horizons, we have the total amount of energy crossing the apparent horizon during $dt$ as
\be
-dE=4\pi R_{A}^2\rho_{tot}(1+w_{tot})dt=3(1+w_{tot})/2dt.
\ee
where $\rho_{tot}=\rho_mf+\rho_{tac}$. Then using (\ref{w}), we get
\be
-dE=(\frac{3}{2}-\frac{\Omega_{tac}}{2}-\frac{\Omega^{3/2}_{tac}}{c}-\frac{4 \pi Q}{3H^{3}})dt.\label{energy}
\ee
On the other hand, with the temperature and holographic
entropy on the apparent horizon we obtain
\bea\label{temp-entro}
T_{A}dS_{A}&=&(\frac{3}{2}-\frac{\Omega_{tac}}{2}-\frac{\Omega^{3/2}_{tac}}{c}-\frac{4 \pi Q}{3H^{3}})dt.
\eea
Thus, we see that the first law of thermodynamics is satisfied on the apparent horizon, i.e., $-dE=T_{A}dS_{A}$.

A similar argument can be applied to the case of event horizon. The energy flow through the event horizon
reads
\bea
-dE&=&4\pi R_{E}^2\rho_{tot}(1+w_{tot})dt=12 \pi c^2\Omega_{tac}^{-1}(1+w_{tot})dt\nonumber\\
&=&12 \pi c^2\Omega_{tac}^{-1}(1-\frac{\Omega_{tac}}{3}-\frac{2 \Omega^{3/2}_{tac}}{3 c}-\frac{8 \pi Q}{9H^{3}})dt.
\eea
To write the first thermodynamics law on event horizon, we need to compute $T_EdS_E =\dot{R_E} dt$ where
\bea\label{redot}
   \dot{R_E}&=&\sqrt{8 \pi}(\frac{c}{\sqrt{\Omega_{tac}}}-1).
\eea
From the above relations it is obvious that $-dE\neq T_{E}dS_{E}$ which means the first law of thermodynamics is violated on the event horizon.

\section{ENTROPY AND SECOND LAW OF THERMODYNAMICS}
We next study the second thermodynamics law and entropy for the model.
From Gibbs relation, the entropy of the universe inside the horizon can be related to its energy and pressure in the
horizon by
\be
T_XdS_{i, X}=dE+p_{tot}dV.
\ee
where $p_{tot}=p_mf+p_{tac}$.

\begin{figure}[h]
\centering
\includegraphics[width=0.4\textwidth]{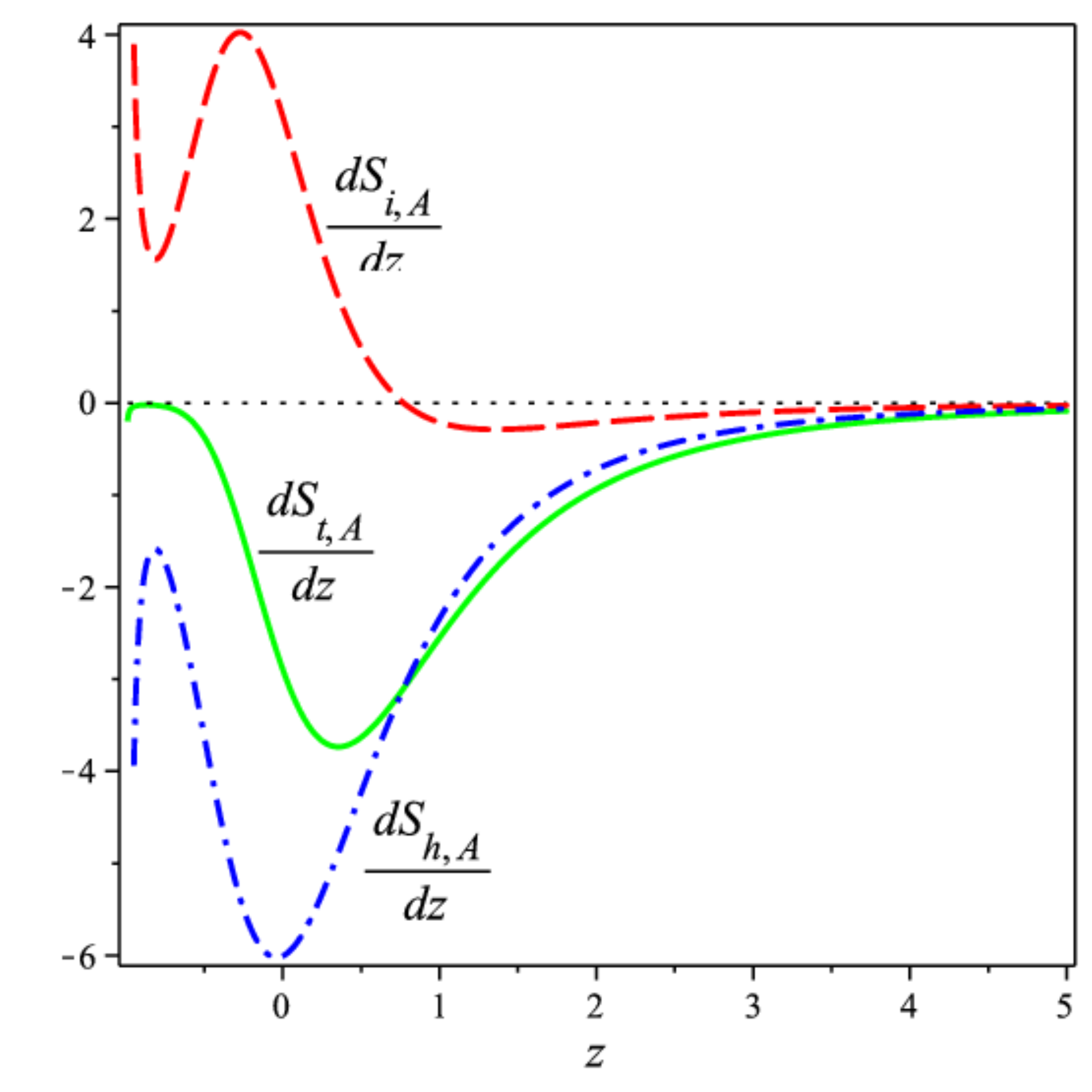}
\includegraphics[width=0.4\textwidth]{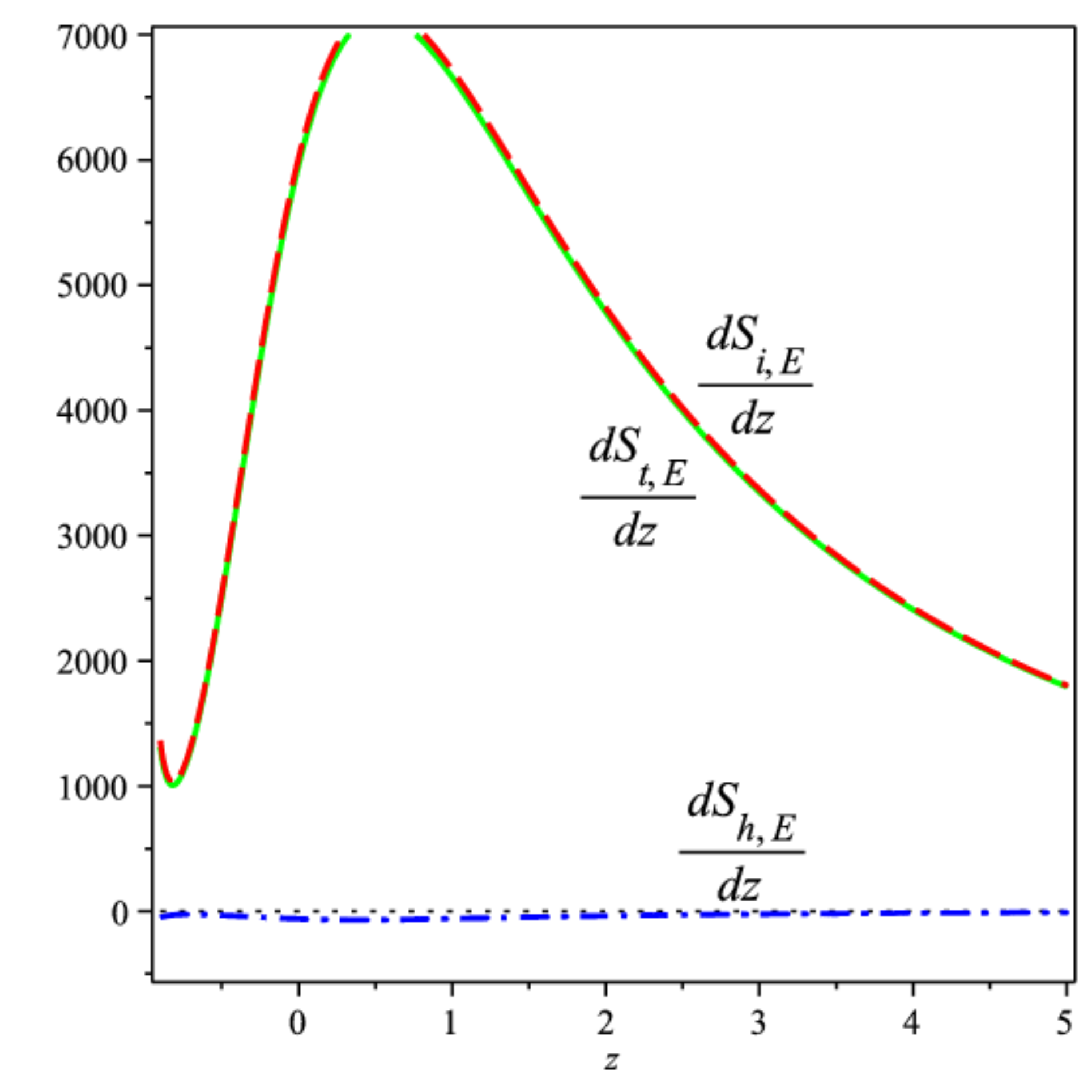}
\caption{The rate of changes of internal entropy $S_{i}$ (red dashed curve), horizon entropy $S_h$ (blue dot-dashed curve) and total entropy $S_t$ (green solid curve) with respect to redshift $z$ using the apparent horizon $R_A$ (left graph) and the event horizon $R_E$ (right graph). We see that GSL is satisfied only in the case of apparent horizon in our model. For plotting we have used the best fit values.}\label{entropy}
\end{figure}

For the apparent horizon, considering the volume $V=4\pi R_{A}^3/3$, energy $E=4\pi \rho_{tot} R_{A}^3/3=R_{A}/2$ and pressure $ p_{tot}=w_{tot}\rho_{tot}=3 w_{tot}/(8\pi R_{A}^2)$, we have
\be
T_AdS_{i, A}=dR_{A}/2+(3/2) w_{tot}dR_{A}.
\ee
Using temperature relation $T_A=1/(2\pi R_{A})$, we then find
\be\label{ds}
dS_{i, A}=\pi (1+3 w_{tot})R_{A}dR_{A}.
\ee
By variation of the internal entropy (\ref{ds}) with respect to redshift, and using the relation $R_{A}dR_{A}=R_{A}\frac{dR_{A}}{dz}dz=-\frac{1}{H^3}\frac{dH}{dz}dz$,
we obtain a differential equation for the entropy enveloped by the apparent horizon given by
\bea\label{dsdtin}
\frac{dS_{i, A}}{dz}&=&-\pi (1+3 w_{tot})H^{-3}\frac{dH}{dz}\nonumber \\
&=&-\pi (1-\Omega_{tac}-\frac{2\Omega_{tac}^{\frac{3}{2}}}{c}-\frac{8 \pi Q}{3H^3})H^{-3}\frac{dH}{dz} \nonumber \\
&=&-2\pi qH^{-3}\frac{dH}{dz}.
\eea
On the other hand, the evolution of the geometric entropy on the apparent
 horizon reads
\be\label{dsdton}
\frac{dS_{h, A}}{dz}=-2 \pi H^{-3}\frac{dH}{dz}\cdot
\ee
From equations (\ref{dsdtin}) and (\ref{dsdton}), a relation between entropy inside and on the apparent horizon can be found as
\bea\label{ration}
\frac{dS_{i, A}}{dz}=q\frac{dS_{h, A}}{dz}
\eea
The above relation shows that for an accelerating universe, $q<0$, the variation of internal and apparent entropy differs by a minus sign. Also, for decelerating universe with $q=1$, their variation with respect to redshift is the same. Finally, one can obtain a differential equation for total entropy as
\be\label{aptoen}
\frac{dS_{t, A}}{dz}=\frac{d(S_{i, A}+S_{h, A})}{dz},
\ee
and test the generalized second law (GSL) in the model. The GSL implies that the total entropy is an increasing function of time, $\dot S_{t, A}\geq0$, or equivalently $\frac{dS_{t, A}}{dz}\leq0$. Fig. \ref{entropy}-left) shows that independent of the dynamics of $\frac{dS_{i, A}}{dz}$ and $\frac{dS_{h, A}}{dz}$, variation of total entropy is always negative. The graph also illustrates that at about $z<0.6$ in the past where universe acceleration begins, $\frac{dS_{i, A}}{dz}>0$ while $\frac{dS_{h, A}}{dz}<0$.

Similarly, one can obtain the entropy inside an event horizon from:
\bea\label{dsdtin2}
\frac{dS_{i, E}}{dz}=\frac{c^4(8\pi)^2}{\Omega_{tac}^2}(-2\pi qH^{-3}\frac{dH}{dz}-\pi(1+q)\Omega_{tac}^{-1}H^{-2}\frac{d\Omega_{tac}}{dz}).
\eea
Also, geometric entropy on the event horizon can be read from
\be\label{dsdton2}
\frac{dS_{h, E}}{dz}=\frac{c^2(8\pi)}{\Omega_{tac}}(-2 \pi H^{-3}\frac{dH}{dz}-\pi \Omega_{tac}^{-1}H^{-2}\frac{d\Omega_{tac}}{dz}).
\ee
From equations (\ref{dsdtin2}) and (\ref{dsdton2}), a relation between entropy inside and on the event horizon can be found as
\begin{equation}\label{relation2}
    \frac{dS_{i, E}}{dz}=\frac{c^2(8\pi)}{\Omega_{tac}}[(1+q)\frac{dS_{h, E}}{dz}+16\pi^2c^2\Omega_{tac}^{-1}H^{-3}\frac{dH}{dz}].
\end{equation}
We therefore derive total entropy variation as
\be\label{evtoen}
\frac{dS_{t, E}}{dz}=\frac{d(S_{i, E}+S_{h, E})}{dz}\cdot
\ee
With numerical calculation Fig. \ref{entropy}-right) shows that $\frac{dS_{t, E}}{dz}>0$ and therefore GSL is not satisfied in this case, irrespective of the universe being in accelerating or decelerating phase.

\section{Summary and Remarks}

The first and generalized second thermodynamics laws are discussed for interacting holographic dark energy model of tachyon cosmology. The interaction term in the formalism is represented by coupling the tachyon scalar field to the matter Lagrangian in the action. The theoretical ground of the cosmological model is tested by observational data. In comparison to $\Lambda$CDM model and for only OHD data our model is not in goof fit. However, when including BAO and CMB data to the dataset, then our model better fits the observational data. By assuming apparent horizon for the universe, the verified model with the observational data does satisfy the first and generalized second thermodynamics laws and also predict recent cosmic acceleration. Note that in our formalism we assume that the scalar field interact only with the matter field. However, one may also consider a geometric coupling of the field same as in Brans-Dicke cosmology or other cosmological models and obtain different results.

\section{appendix}

We constrain the parameters including the initial conditions by minimizing the $\chi^2$ function given as
\begin{widetext}
\begin{equation}\label{chi2}
    \chi^2_{OHD}(c ,b; \Omega_{tac}(0), H(0), \phi(0))=\sum_{i=1}^{14}\frac{[H^{th}(z_i|c ,b; \Omega_{tac}(0), H(0), \phi(0)) - H^{obs}(z_i)]^2}{\sigma_{OHD}^2(z_i)},
\end{equation}
\end{widetext}
where the sum is over the cosmological dataset. In (\ref{chi2}),  $H^{th}$ and $H^{obs}$ are the Hubble parameters obtained from the theoretical model and from observation, respectively. Also, $\sigma_{OHD}$ is the estimated error of the $H^{obs}$ where obtained from observation \cite{Cao}.

We add the CMB data in our analysis. Since the CMB shift parameter $R$ \cite{r1},\cite{r2}, contains the main information of the observations from the CMB, it is used to constrain the theoretical models by minimizing
\begin{equation}\label{chicmb}
    \chi^2_{CMB}=\frac{[R-R_{obs}]^2}{\sigma_R^2},
\end{equation}
where $R_{obs} = 1.725\pm0.018$ \cite{r3}, is given by the WMAP7 data. Its corresponding theoretical value is defined as
\begin{equation}\label{r}
    R\equiv\Omega_{m0}^{1/2}\int_0^{z_{CMB}}\frac{dz'}{E(z')},
\end{equation}
with $z_{CMB} = 1091.3$.

Moreover, for the BAO data, the BAO distance ratio at $z = 0.20$ and $z = 0.35$ from the joint analysis of the 2dF Galaxy Redsihft Survey and SDSS data \cite{snss}-\cite{He}  is used. The distance ratio, given by
\begin{equation}\label{drbao}
   \frac{D_V(z=0.35)}{D_V(z=0.20)}=1.736\pm0.065,
\end{equation}
is a relatively model independent quantity with $D_V(z)$ defined as
\begin{equation}\label{bao}
    D_V(z_{BAO})=[\frac{z_{BAO}}{H(z_{BAO})}(\int_0^{z_{BAO}}\frac{dz}{H(z)})^2]^{1/3}.
\end{equation}
So, the constraint from BAO can be obtained by performing the following $\chi^2$ statistics
\begin{equation}\label{chibao}
    \chi^2_{BAO}=\frac{[(D_V(z=0.35)/D_V(z=0.20))-1.736]^2}{0.065^2}\cdot
\end{equation}
The constraints from a combination of OHD, BAO and CMB can be obtained by minimizing $\chi^2_{OHD}+\chi^2_{BAO}+\chi^2_{CMB}$.

\section{Acknowledgement}
This work is supported in part by Research Grant Council of University of Guilan.


\begin{thebibliography}{99}

\bibitem{faraj} H. Farajollahi, A. Ravanpak and G. Farrpoor Fadakar, Mod. Phys. Lett. A. 26, 15, 1125-1135 (2011).
\bibitem{Farajj} H. Farajollahi, A. Ravanpak and G. F. Fadakar, Phys. Lett. B. 711, 3-4, 15, 225-231 (2012).
\bibitem{faraj1} H. Farajollahi and A. Salehi, Phys. Rev. D. 83, 124042 (2011).
\bibitem{faraj2} H. Farajollahi, A. Salehi and A. Shahabi, JCAP. 10, 014 (2011).
\bibitem{faraj3} H. Farajollahi, A. Salehi, F. Tayebi and A. Ravanpak, JCAP. 05, 017 (2011).
\bibitem{Ravanpak} H. Farajollahi, A. Ravanpak and G. F. Fadakar, Astrophys. Space. Sci. 336, 2, 461-467 (2011).
\bibitem{Riess} A. G. Riess, et al., Astron. J. 116, 1009 (1998).
\bibitem{Perlmutter} S. Perlmutter, et al., Astrophys. J. 517, 565 (1999).
\bibitem{Spergel} D. N. Spergel, et al., ApJS, 148, 175 (2003).
\bibitem{Spergel2} D. N. Spergel, et al., ApJS, 170, 377S (2007).
\bibitem{Tegmark} M. Tegmark, et al., Phys. Rev. D. 69, 103501 (2004).
\bibitem{Eisenstein} D. J. Eisenstein, et al., Astrophys. J. 633, 560 (2005).
\bibitem{Caldwell} R. R. Caldwell,  R. Dave and  R. J. Steinhardt, Phys. Rev. Lett. 80, 1582 (1998).
\bibitem{Caldwell2} R. R. Caldwell, Phys. Lett. B. 545, 23 (2002).
\bibitem{Armendariz} C. Armendariz-Picon,  V. Mukhanov and  P. J. Steinhardt, Phys. Rev. D. 63, 103510 (2001).
\bibitem{Padmanabhan2} T. Padmanabhan, Phys. Rev. D. 66, 021301 (2002).
\bibitem{Sen} A. Sen, Phys. Scripta. T. 117, 70 (2005).
\bibitem{Feng} B. Feng, X. L. Wang and  X. M. Zhang, Phys. Lett. B. 607, 35 (2005).
\bibitem{eli} E. Elizadle, S. Nojiri and  S. D. Odintsov, Phys. Rev. D. 70, 043539 (2004).
\bibitem{Kamenshchik} A. Kamenshchik, U. Moschella and V. Pasquier, Phys. Lett. B. 511, 265 (2001).
\bibitem{Bento} M. C. Bento, O. Bertolami and A. A. Sen, Phys. Rev. D. 66, 043507 (2002).
\bibitem{Cohen} A. G. Cohen, D. B. Kaplan and A. E. Nelson, Phys. Rev. Lett. 82, 4971 (1999).
\bibitem{Li} M. Li, Phys. Lett. B. 603, 1 (2004).
\bibitem{Wei} H. Wei and R. G. Cai, Phys. Lett. B. 663, 1 (2008).
\bibitem{Wei2} H. Wei and R. G. Cai, Phys. Lett. B. 660, 113 (2008).
\bibitem{Gao} C. Gao, F. Wu, X. Chen and Y. G. Shen, Phys. Rev. D. 79, 043511 (2009).
\bibitem{Hsu} S. D. H. Hsu, Phys. Lett. B. 594, 13 (2004).
\bibitem{Huang} Q. G. Huang and M. Li, JCAP. 0408, 013 (2004).
\bibitem{Wang} B. Wang, Y. g. Gong and E. Abdalla, Phys. Rev.  D 74, 083520 (2006).
\bibitem{Bin} B. Wang, J. Zang, C. Y. Lin, E. Abdalla and S. Micheletti, Nucl. Phys. B. 778, 69-84 (2007).
\bibitem{Amen} L. Amendola, Phys. Rev. D. 62, 043511 (2000).
\bibitem{Sand} H. Sandvik, M. Tegmark, M. Zaldarriaga and I. Waga, Phys. Rev. D. 69, 123524 (2004).
\bibitem{mohseni2}  H. Mohseni Sadjadi and M. Honardoost, Phys. Lett. B. 647, 231 (2007).
\bibitem{Q} Q. Wu, Y. Gong, A. Wang and J. S. Alcaniz, Phys. Lett. B. 659, 34 (2008).
\bibitem{Brax} P. Brax, C. V. Bruck, D. F. Mota, N. J. Nunes and H. A. Winther, Phys. Rev. D. 82, 083503 (2010).
\bibitem{davis} A. C. Davis, C. A. O. Schelpe and D. J. Shaw, Phys. Rev. D. 80, 064016 (2009).
\bibitem{ito} Y. Ito and S. Nojiri, Phys. Rev. D. 79, 103008 (2009).
\bibitem{wung} W. H. Huang, Phys. Lett. B. 561, 153 (2003).
\bibitem{Bagla} J. S. Bagla, H. K. Jassal and T. Padmanabhan, Phys. Rev. D. 67, 063504, (2003).
\bibitem{Setare3} M. R. Setare, Phys. Lett. B. 653, 116-121 (2007).
\bibitem{Setare5} M. R. Setare, jcap. 0701, 023 (2007).
\bibitem{carter} J. M. Bardeen, B. Carter and S. W. Hawking, Comm. Math. Phys. 31, 161 (1973).
\bibitem{Hawking} S. W. Hawking, Comm. Math. Phys. 43, 199 (1975).
\bibitem{Bekenstein} J. D. Bekenstein, Phys. Rev. D. 7, 2333 (1973).
\bibitem{Jacobson} T. Jacobson, Phys. Rev. Lett. 75, 1260 (1995).
\bibitem{Verlinde} E. Verlinde, (hep-th/0008140).
\bibitem{Padmanabhan} T. Padmanabhan, Class. Quantum. Grav. 19, 5387 (2002).
\bibitem{Cai} R. G. Cai and S. P. Kim, JHEP. 02, 050 (2005).

\bibitem{Bhattacharya} S. Bhattacharya and U. Debnath, Int. J. Theor. Phys. 50, 525-536 (2011).

\bibitem{Bamba} K. Bamba, C. Q. Geng and S. Tsujikawa, Int. J. Mod. Phys. D. 20, 1363-1371 (2011).
\bibitem{Zhang} Y. Zhang, Y. G. Gong and Z. H. Zhu, Int. J. Mod. Phys. D. 20, 8, 1505-1519 (2011).
\bibitem{Bamba2} K. Bamba, C. Q. Geng and S. Tsujikawa, Phys. Lett. B. 688, 101-109 (2010).

\bibitem{Chattopadhyay} S. Chattopadhyay and U. Debnath, Can. J. Phys. 88, 933-938 (2010).

\bibitem{Mazumder} N. Mazumder and S. Chakraborty, Gen. Rel. Grav. 42, 813-820 (2010).
\bibitem{Setare} M. R. Setare and E. C. Vagenas, Phys. Lett. B. 666, 111-115 (2008).
\bibitem{Wang2} B. Wang, C. Y. Lin, D. Pavon and E. Abdalla, Phys. Lett. B. 662, 1-6 (2008).
\bibitem{Setare2} M. R. Setare, JCAP. 0701, 023 (2007).

\bibitem{Cao} S. Cao, Z. H. Zhu and N. Liang, Astronomy and Astrophysics. 529, A61 (2011).
\bibitem{r1} Y. Wang and P. Mukherjee, Astrophys. J. 650, 1 (2006).
\bibitem{r2} J. R. Bond, G. Efstathiou and M. Tegmark, Mon. Not. Roy. Astron. Soc. 291, L33 (1997).
\bibitem{r3} E. Komatsu et al., Astrophys. J. Suppl. 192, 18 (2011).
\bibitem{snss} B. A. Reid, Mon. Not. Roy. Astron. Soc. 404, 60-85 (2010).
\bibitem{Perc} W. J. Percival, et al., Mon. Not. Roy. Astron. Soc. 401, 2148-2168 (2010).
\bibitem{He} J. H. He, B. Wang and P. Zhang, Phys. Rev. D. 80, 063530 (2009).


\end{thebibliography}
\end{document}